\documentclass[final]{cvpr}

\usepackage{times}
\usepackage{epsfig}
\usepackage{graphicx}
\usepackage{amsmath}
\usepackage{amssymb}
\usepackage{epstopdf}
\usepackage{subfigure}
\usepackage{multirow}
\usepackage{amsthm}
\usepackage{bm}
\usepackage{makecell}
\usepackage{color}
\usepackage{booktabs}
\usepackage{caption}
\usepackage[ruled,linesnumbered]{algorithm2e}
\usepackage[pagebackref=true,breaklinks=true,colorlinks,bookmarks=false]{hyperref}

\newcommand{\CF}{\mathcal{F}}

\newcommand{\HC}{\hat {\mathbf{C}}}
\newcommand{\Tz}{\tilde{z}}

\newcommand{\Bx}{\mathbf{x}}
\newcommand{\BZ}{\mathbf{Z}}
\newcommand{\BR}{\mathbf{R}}
\newcommand{\hc}{\hat{c}}
\newcommand{\Hm}{\hat{m}}

\newcommand{\ta}{\theta}

\newcommand{\Bu}{\mathbf{u}}
\newcommand{\Bv}{\mathbf{v}}

\newcommand{\by}{\mathbf{y}}

\newcommand{\BS}{\mathbf{S}}
\newcommand{\BAT}{\breve{\mathbf{\Theta}}}

\newcommand{\BK}{\mathbf{K}}
\newcommand{\BI}{\mathbf{I}}
\newcommand{\Be}{\mathbf{e}}

\newcommand{\BU}{\mathbf{U}}
\newcommand{\BV}{\mathbf{V}}
\newcommand{\BC}{\mathbf{C}}

\newcommand{\BE}{\hat{\mathbf{E}}}
\newcommand{\HW}{\hat{\mathbf{W}}}
\newcommand{\BT}{\mathbf{\Theta}}

\newcommand{\BP}{\mathbf{P}}

\newcommand{\tz}{\tilde{z}}
\newcommand{\CL}{\mathcal{L}}
\newcommand{\TT}{\tilde{\mathbf{\Theta}}}
\newcommand{\HT}{\hat{\mathbf{\Theta}}}

\newtheorem{theorem}{$\mathbf{Theorem}$}

\def\cvprPaperID{3239} 
\def\confYear{CVPR 2021}

\begin{document}

\title{Probabilistic Selective Encryption of \\ Convolutional Neural Networks for Hierarchical Services}

\author{Jinyu Tian\textsuperscript{\rm 1}, Jiantao Zhou\textsuperscript{\rm 1,*}, and Jia Duan\textsuperscript{\rm 2}
\\
\textsuperscript{\rm 1}State Key Laboratory of Internet of Things for Smart City, \\ Department of Computer and Information Science, University of
Macau\\
\textsuperscript{\rm 2}JD Explore, JD\\
{\tt\small \{yb77405, jtzhou\}@um.edu.mo, duanjia1@jd.com}
}

\maketitle

\begin{abstract}
   Model protection is vital when deploying Convolutional Neural Networks (CNNs) for commercial services, due to the massive costs of training them. In this work, we propose a selective encryption (SE) algorithm to protect CNN models from unauthorized access, with a unique feature of providing hierarchical services to users.  Our algorithm firstly selects important model parameters via the proposed Probabilistic Selection Strategy (PSS). It then encrypts the most important parameters with the designed encryption method called Distribution Preserving Random Mask (DPRM), so as to maximize the performance degradation by encrypting only a very small portion of model parameters. We also design a set of access permissions, using which different amount of most important model parameters can be decrypted. Hence, different levels of model performance can be naturally provided for users. Experimental results demonstrate that the proposed scheme could effectively protect the classification model VGG19 by merely encrypting $8\%$ parameters of convolutional layers. We also implement the proposed model protection scheme in the denoising model DnCNN, showcasing the hierarchical denoising services.
   
 \end{abstract}


\begin{figure*}[!th]
	\centering
	\includegraphics[width = 0.8 \textwidth,height=0.21 \textwidth]{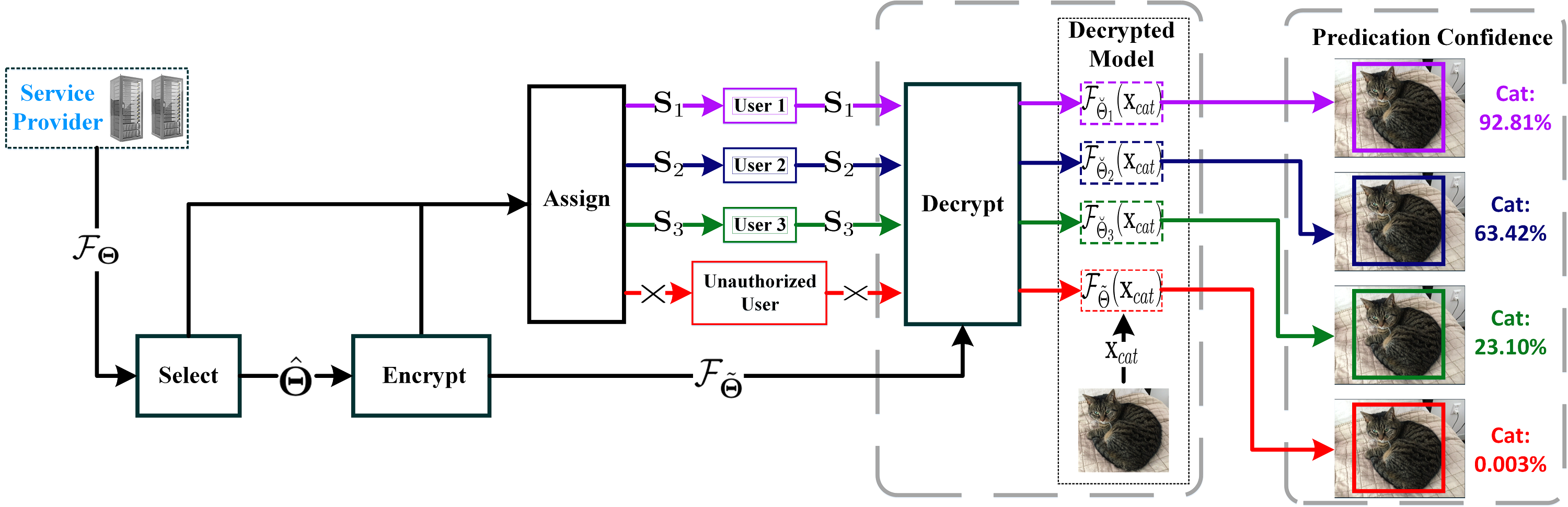}
	\caption{\small{Schematic diagram of our proposed system model.}}  \label{Fig:Framework}
	\vspace*{-0.4cm}
\end{figure*}

\section{Introduction}
Convolutional Neural Network (CNN) has been used to achieve unparalleled results across various tasks such as object detection \cite{Obj1,Obj2,Obj3}, super-resolution reconstruction \cite{Recons3,Recons1,Recons2}, and image inpainting \cite{inpainting3,inpainting1,inpainting2}. The construction of a successful CNN model is not a trivial task, which usually requires substantial investments in expertise, time, and resources. To encourage healthy business investment and competitions, it is crucial to prevent unauthorized access to CNN models. Meanwhile, there is a recent trend of deploying pretrained CNN models through cloud-based services. Under such circumstance, it is much desirable to offer hierarchical services, such that users with different access privileges could enjoy different levels of model performance. For instance, when using CNN model for image denoising task, the lowest access privilege leads to a roughly denoised version; this could serve as a promotion of the denoising service, which could be free. When users prefer sophisticatedly denoised images, they can pay for advanced access privileges for better services.

A straightforward strategy to achieve model protection and access control is to encrypt all model parameters via traditional cryptographic methods such as RSA \cite{RSA}, TDES \cite{TDES}, and Twofish  \cite{Twofish}. On the order of millions or more parameters, both the encryption and decryption of them would be very expensive, especially for resource-constrained devices. In comparison, it is much desirable to encrypt parameters selectively. This strategy is known as Selective Encryption (SE) and has practical applications in multimedia security and Internet security \cite{SE2,25,Marco,24,SE4}. On the other hand, encrypting all model parameters indiscriminately cannot provide fine-grained hierarchical services.

The motivation of the SE derives from Shannon's seminar work, which pointed out that effective encryption/decryption can be performed by decreasing the redundancy of a system \cite{shannon}. In general, reducing the redundancy of a system relies on the prior knowledge of the importance of system components.  For instance, Abomhara \emph{et. al.} \cite{Abomhara} protected the bitstreams of H.264/AVC  video codec by selectively encrypting the I-frames, which have larger impacts on the quality of reconstructed frames, compared with P- and B- frames. In fact, in the deep learning community, several works \cite{importance1,importance2,importance4,importance3} showed that the parameters in a CNN model are \emph{NOT} equally important, and some of them contain more useful information for a given task. The unequal importance of model parameters motivates us to design a new SE for protecting a CNN model by only encrypting those important parameters.

Therefore, in this work, we present a novel SE algorithm to protect a pretrained CNN model, with a unique feature of providing hierarchical services to users. Our proposed protection scheme consists of the following two steps. First of all, we propose the \emph{Probabilistic Selection Strategy} (PSS) to select important model parameters, which heavily impact the model performance. Then, we design an encryption algorithm \emph{Distribution Preserving Random Mask} (DPRM) for encrypting those selected parameters. As will be verified experimentally, the proposed SE endows the feasibility of providing users with hierarchical services, i.e., different levels of model performance could be granted according to predefined permissions. The main contributions of our work can be summarized as follows:

\begin{itemize}
	\item We propose the PSS to determine the importance of parameters in a CNN model. The PSS could be generalized to CNN models for different applications, such as image classification and denoising.
	
	\item We propose the algorithm DPRM to encrypt parameters selected by the PSS such that encrypted parameters and those unencrypted ones are statistically consistent. We theoretically prove that the ciphertext obtained from the DPRM is imperceptible to attackers, which significantly enhances the security of the protected model.
	
	\item Through manipulating the number of decrypted parameters with different level of importance, the proposed framework could provide hierarchical services for users by assigning different permissions.
	
	\item The experimental results on the classification model VGG19 and the denoising model DnCNN show that we can effectively protect the two models with the proposed SE, by merely encrypting less than 8\% parameters of convolutional layers.
\end{itemize}

The rest of the paper is organized as follows. Section \ref{Sec:Framemwork} briefly introduces the system model, the threat model, and design goals of the proposed framework. We provide the details about the system model in Section \ref{Sec:Details}. In Section \ref{Sec:Imperceptibility}, we theoretically prove that the ciphertext of those important parameters is imperceptible to attackers. Finally, Section \ref{Sec:Exps} offers experimental results to verify that all expected goals are achieved, and Section \ref{Sec:Conclusion} concludes. 

\section{System model, threat model and design goals}\label{Sec:Framemwork}

\subsection{System model} \label{Sec:SystemModel}
We aim to design a CNN protection framework with the following two functionalities. Firstly, this system can prevent unauthorized access to the pretrained model. Namely, only authorized users could obtain the correct model outputs, while the unauthorized ones can only get irrelevant results. Secondly, it can provide hierarchical services for authorized users with different permissions. The framework of our system is presented in Fig. \ref{Fig:Framework}. Let $\CF_{\BT}$ be a pretrained CNN model with the parameter set $\BT$. To protect this model via our proposed SE strategy, a service provider firstly feeds the pretrained model $\CF_{\BT}$ into a {\bf{Select}} module, which selects important parameters $\HT$ from $\CF_{\BT}$ by using the proposed PSS.  Then, the {\bf{Encrypt}} module constructs the protected model $\CF_{\TT}$ by encrypting $\HT$ with the proposed DPRM. The model $\CF_{\TT}$ subsequently will be deployed, waiting for access with permissions. The {\bf{Assign}} module generates several permissions $\BS_{\Hm}$'s based on the outputs of the {\bf{Select}} and {\bf{Encrypt}} modules. Authorized users then obtain a decrypted model  $\CF_{\BAT_{\Hm}}$ by inputting a permission $\BS_{\Hm}$ into the {\bf{Decrypt}} module. The decrypted model will exhibit different levels of performance of the pretrained model $\CF_{\BT}$ as the change of the permission $\BS_{\Hm}$. As shown in Fig. \ref{Fig:Framework}, three authorized users decrypt the protected model with different permissions $\BS_{\Hm}$'s ($\Hm=1,2,3$), and obtain corresponding decrypted models $\CF_{\BAT_{\Hm}}$ with hierarchical performance on classifying a cat. When an unauthorized user attempts to access the protected model without any permission, the system will return a useless result.

\subsection{Threat model} \label{Sec:Threat}

The considered security threats of the proposed system mainly come from attackers who attempt to recover encrypted parameters $\HT$ from the protected model $\CF_{\TT}$, so as to use the pretrained model $\CF_{\BT}$ without authorization.  On the one hand, attackers could treat the parameter set $\TT$ of the protected model as a noisy version of the original parameter set $\BT$, where parameters $\HT$ are contaminated. Attackers thus can recover $\HT$ from $\TT$ with denoising techniques. On the other hand, those encrypted parameters $\HT$ could be treated as the missing information in $\TT$, which possibly can be restored with retraining strategies. We call these two adversarial behaviors the \emph{denoising attack} and the \emph{retraining attack}.

\subsection{Design goals}\label{Sec:goals}

To evaluate the effectiveness and security of the proposed scheme, we clarify the following design goals: 1) {\bf{Effectiveness}}: To make the protected model $\CF_{\TT}$  dysfunctional to unauthorized access, we have to ensure that the selectively encrypted model exhibits sufficiently large performance degradation; 2) {\bf{Hierarchy}}: The proposed system should provide users with different services by assigning them with different permissions. Therefore, the released model $\CF_{\BAT_{\Hm}}$ needs to exhibit various performance in accordance to user's permission $\BS_{\Hm}$; and 3) {\bf{Security}}: The protected model $\CF_{\TT}$ should be secure against threats discussed above.  We will verify all the design goals in Section \ref{Sec:Exps}.

\begin{figure*}[!t]
	\centering
	\subfigure{
		{\includegraphics[width=0.23\linewidth]{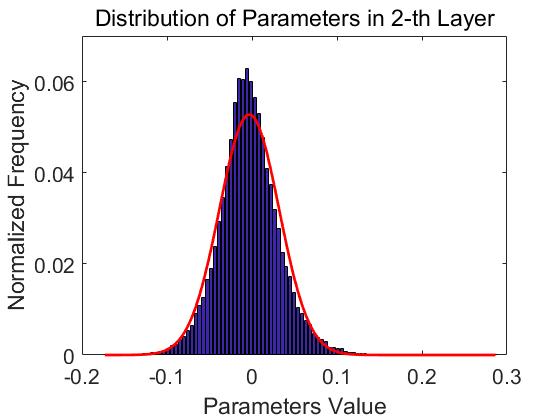}}	
	}
	\subfigure{	
		{\includegraphics[width=0.23\linewidth]{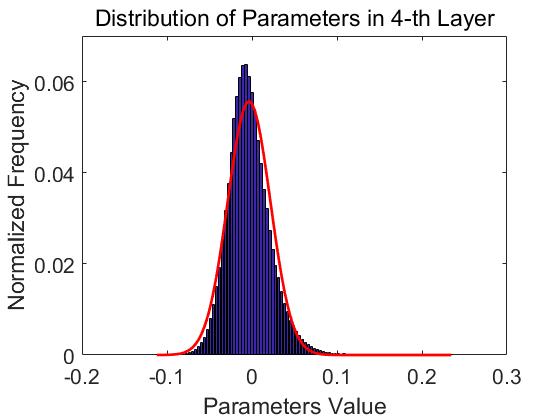}}	
	}
	\subfigure{
		{\includegraphics[width=0.23\linewidth]{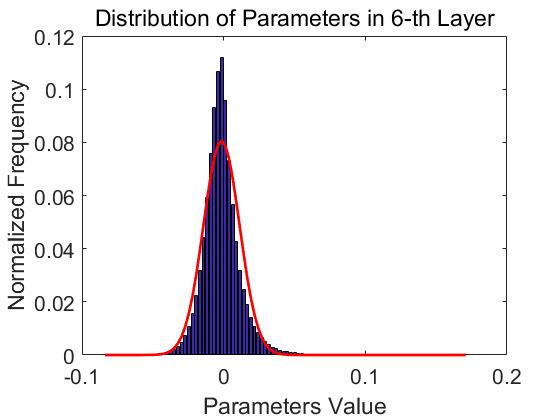}}		
	}
	\subfigure{
		{\includegraphics[width=0.23\linewidth]{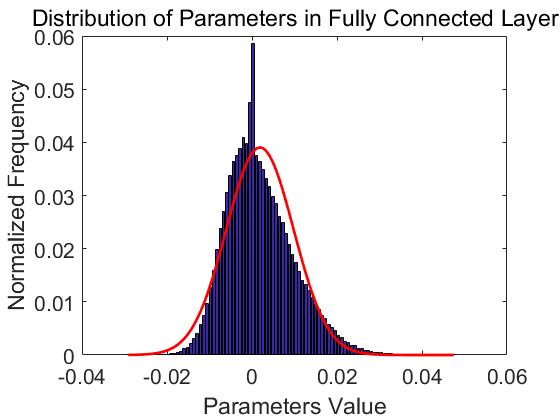}}	
	}
	\caption{\small{The histograms of parameters in several layers of VGG19. The red curve in each sub-figure is the corresponding Gaussian distribution with mean value and standard deviation estimated from parameters.}} \label{Fig:hist}
	\vspace*{-0.4cm}
\end{figure*}

\section{The system model for protecting CNN}\label{Sec:Details}
In this section, we provide details concerning each module of our proposed system model in Section \ref{Sec:SystemModel}.

\subsection{Select} \label{Select}

This module aims to select important parameters $\HT$ from convolutional layers of $\CF_{\BT}$. Here we only consider convolutional layers because most of parameters in a CNN model are concentrated in these layers \cite{res,Lenet5,GoogleNet,inc}. Hereafter, the term ``layer" refers to the convolutional layer. We adopt a layer-wise strategy to determine $\HT$ for parallel processing. Specifically, suppose the pretrained model contains $L$ layers, and hence, $\HT$ is composed of $L$ subsets $\HT^l$'s ($l = 1,...,L$), where each $\HT^l$ contains important parameters of the $l$-th layer. We propose a selection method PSS to identify each $\HT^l$. The motivation of PSS is as follows.

Intuitively, we can identify the importance of parameters by evaluating the performance degradation of the pretrained model without these parameters. However, neurons of a CNN model may have different responses for various inputs, implying that the importance of parameters is related to the inputs. More precisely, let $\BT^{l}$ denote parameters of the $l$-th layer. When feeding a sample $\Bx_n$ ($n=1,...,N$) into  $\CF_{\BT}$, there exists a parameter subset $\HT^{l}$ of $\BT^l$ to cause the maximal performance degradation of $\CF_{\BT}$ if we remove $\HT^{l}$. Clearly, $\HT^{l}$ changes with the input $\Bx_n$. To eliminate such randomness, we can count how many times each parameter $\theta$ in $\BT^l$ is selected as a candidate of $\HT^l$, after feeding all $\Bx_n$. Denote the selected frequency of a parameter $\theta$ in $\BT^l$ by $p_{\theta}$. It is clear that the frequency $p_{\theta}$ directly reflects the importance of a parameter $\theta$ to the pretrained model. We thus name $p_{\ta}$ the importance of the parameter $\ta$. For simplicity, we call $\HT^l$ \emph{dominated set} of the $l$-th layer and call parameters in it \emph{dominated parameters}. Naturally, $\HT$ is the dominated set of $\CF_{\BT}$.
 
Now we formulate the selection strategy PSS above into the following optimization problem.
\begin{equation} \label{OPT}
\begin{small}
	\min_{p_\ta\in[0,1]} \frac{1}{N}\sum_{n=1}^{N} \CL(\CF_{\BT}(\Bx_n,(\BI-\BZ^{(n)}) \odot \BT^l),\by_n) + \lambda \|\BZ^{(n)}\|_0,
\end{small}
\end{equation}
where $\BZ^{(n)} = \{z^{(n)}_{\ta}\}_{\ta \in \BT^l}$, $z^{(n)}_{\ta} \sim Bern(z_\ta|p_\ta)$ is a sample of the binary random variable $z_\ta$, $\BI$ is the vector of ones with the same length as $\BT^l$, and $\lambda$ is a weighting factor for the regularization term. Here, we briefly explain the relationship between the optimization problem (\ref{OPT}) and the selection strategy. The element-wise multiplication $(\BI-\BZ^{(n)}) \odot \BT^l$ is designed to simulate the removing operation by noting that a parameter $\ta$ will be removed from $\BT^l$ if the corresponding $z^{(n)}_\ta=1$. The first term thus indicates the performance of $\CF_{\BT}$ after removing a part of the parameters ($\CL(\cdot)$ is a performance evaluating function). To maximize the performance degradation of $\CF_{\BT}$, which is equivalent to minimizing the first term in (\ref{OPT}), those important parameters $\ta$'s in $\BT^l$ should be assigned with large importance $p_{\ta}$, so that $z^{(n)}_\ta=1$ for most $\Bx_n$'s. Thus, we can determine the importance $p_\ta$ of each $\ta$ in $\BT^l$ by solving the problem (\ref{OPT}). It should be noted that the term $\|\BZ^{(n)}\|_0$ penalizes the number of removed parameters, such that fewer parameters are assigned with large importance.

The discrete nature of the binary random variable $z_\ta$ challenges the optimization of the problem (\ref{OPT}). We can overcome this obstacle via \emph{reparameterizing} \cite{reparameterization,reparameterization2} $z_\ta$ into the continuous random variable $\tilde{z}_\ta$ as follows
\begin{equation} \label{Concert}
\begin{small}
\begin{aligned}
&s_\ta(u) = Sig((\log (u/(1-u))+\log(p_\ta/(1-p_\ta)))/\beta), \\
&\tilde{s}_\ta(u) = s_\ta(u)(\zeta-\gamma)+\gamma,\\
&\tilde{z}_\ta = \min(1,\max(0,\tilde{s}_\ta(u)),
\end{aligned}
\end{small}
\end{equation}
which was initially proposed in \cite{maddison} and later improved in \cite{Christos}. In (\ref{Concert}), $u$ is a random variable of uniform distribution $U(0,1)$, $Sig(\cdot)$ represents the sigmoid function, $\zeta>0$ and $\gamma<0$ are parameters to extend the support of $\tilde{z}$ to be $[0,1]$.

Another challenge ascribes to the $L_0$ regularizer $\|\tilde{\BZ}^{(n)}\|_0 = \|\{\tz^{(n)}_{\ta}\}_{\ta \in \BT^l}\|_0$ in the problem (\ref{OPT}). Note that the original $z_{\ta}$ has been relaxed by $\tz_{\ta}$ defined in (\ref{Concert}). We can relax the $L_0$ regularizer into a differentiable form {\small{$\sum_{\ta\in\BT^l}\mathbb{P}\left\{ \Tz^{(n)}_{\ta} \neq 0 \right\}$}}. The reason is that $L_0$ norm enforces less nonzero elements in $\tilde{\BZ}^{(n)}$. This implies that, for most $\tz^{(n)}_{\ta}$'s in $\tilde{\BZ}^{(n)}$, the probability {\small{$\mathbb{P}\left\{ \Tz^{(n)}_{\ta} \neq 0 \right\}$}} should be as small as possible. According to the cumulative distribution function (CDF) of $s_\ta(u)$ introduced in \cite{maddison}, we can conclude that
\begin{equation} \label{phz}
\begin{small}
\mathbb{P}\left\{ \tz^{(n)}_\ta \neq 0 \right\} = Sig \left( \log(p_\ta/(1-p_\ta))-\beta \log{\frac{-\gamma}{\zeta}} \right),
\end{small}
\end{equation}
where the details is given in supplementary materials.

Upon having the relaxation (\ref{Concert}) and (\ref{phz}), the problem (\ref{OPT}) reduces to
\begin{equation} \label{OPT2}
\begin{small}
\begin{aligned}
&\min \limits_{p_\ta\in[0,1]} ~ \frac{1}{N} \sum_{n=1}^{N} \left(\CL(\CF_{\BT}(\Bx_n,(\BI-\tilde{\BZ}^{(n)}) \odot \BT^l ),\by_n) \right), \\
&+ \lambda \sum_{\ta \in \BT^l}Sig( \log(p_\ta/(1-p_\ta))-\beta \log{\frac{-\gamma}{\zeta}}).
\end{aligned}
\end{small}
\end{equation}
When the performance evaluation function $\CL(\cdot)$ is differentiable with respect to $p_\ta$'s, such as the negative cross-entropy or negative mean squared error, we can effortlessly solve this problem with automatic differentiation toolboxes TensorFlow \cite{Tensorflow} or Pytorch \cite{pytorch}.

After solving the problem (\ref{OPT2}), we obtain the importance of parameters in the $l$-th layer, denoted by $\BP^l =\{p_\ta\}_{\ta \in \BT^l}$. The dominated set $\HT^l$ of this layer naturally can be identified as follows
\begin{equation} \label{DPl}
\begin{small}
	\HT^l = \{\theta | \theta \in \BT^l, p_{\theta}~\text{is~top}~\phi~\text{in}~\BP^l\},
\end{small}
\end{equation}
where $\phi$ is the number of dominated parameters. By repeatedly solving the problem (\ref{OPT2}) with $l=1,...,L$, and searching dominated parameters as (\ref{DPl}), we can obtain the dominated set $\HT = \{\HT^l\}^L_{l=1}$ for the model $\CF_{\BT}$.

Upon having dominated parameters $\HT$, we now split them into $M$ different subsets as follows:
\begin{equation} \label{partition}
\begin{small}
\begin{aligned}
&\HT = \{\HT_m\}_{m=1}^M, \quad \HT_m = \{ \HT_m^l \}^L_{l=1}, \\
&\HT_m^l = \{\ta|\ta\in\HT^l, Q_{m+1,l} < p_\ta \leq Q_{m,l} \},
\end{aligned}
\end{small}
\end{equation} 	
where $Q_{m,l}$ is the $(M-m+1)/M$ percentile of elements in $\BP^l$.
Here we split $\HT$ according to the importance of parameters as in (\ref{partition}), so that the subsequent procedure could control the performance of $\CF_{\BT}$ by manipulating the number of decrypted parameters with different importance. More details are deferred to Section \ref{Sec:Decrypt}.

For the future decryption of dominated parameters in $\HT$, we also need their locations $\BE=\{\Be_{\ta}\}_{\ta\in\HT}$ in the model $\CF_{\BT}$, where $\Be_{\theta}$ is a 2-D tuple to locate which layer $\ta$ belongs to and the position of $\ta$ in this layer. Similarly, we split $\BE$ following the partition of $\HT$. That is,
\begin{equation}
\begin{small}
\BE = \{\BE_m\}_{m=1}^{M},~\BE_m = \{\BE^l_m\}_{l=1}^{L},~\BE_m^l = \{\Be_{\ta}\}_{\ta\in\HT_m^l}.
\end{small}
\end{equation} 	

\begin{algorithm}[!t]
	\SetAlgoNoEnd
	\footnotesize{
		\caption{The Scheme of the $Encrypt$ module (running the DPRM on each subset of $\HT$).} \label{Alg:Encrypt}
		\KwIn{Dominated set $\HT = \{\HT_{m}\}_{m=1}^M$, secret keys $\BK = \{\kappa_m\}_{m=1}^M$.}
		\KwOut{$\CF_{\TT},\BU,\BV,\BK,\bm{\mu},\bm{\sigma}$.} 		
		\For{$ ~m = 1,...,M$}{
			\mbox{$\BR_{m} \leftarrow \mathcal{G}_{fs}(\kappa_m,|\HT_{m}|), \quad \BC_{m} = \HT_{m}+\BR_{m}$}\;
			\For{$l = 1,...,L$}{
				\mbox{$\HC^l_{m}=\emptyset,~\BC_m=\{\BC^l_{m}\}^L_{l=1}$}\;
				\For{$c ~\mathbf{in}~ \BC^l_{m}$} {
					\mbox{$u^l_m = \min (\BC^l_{m}), \quad v^l_m = \max(\BC^l_{m})$}\;
					\mbox{$c \leftarrow (c-u^l_m)/(v^l_m-u^l_m)$}\;
					\mbox{$\mu^l =  \text{MEAN}(\BT^l), \quad \sigma^l = \text{STD}(\BT^l)$}\;
					\mbox{$\hc \leftarrow F^{-1}\left(c|\mu^l,\sigma^l\right),~\HC^l_{m} \leftarrow \HC^l_{m} \bigcup \hc$}\;
				}
			}
			\mbox{$\HC_{m} = \{\HC^l_{m}\}^L_{l=1}$}\;
		}
		\mbox{$\HC = \{\HC_{m}\}^M_{m=1},~\CF_{\TT} \leftarrow \text{replace}~\HT~\text{in}~\CF_{\BT}	~\text{with}~\HC$}\;
		\mbox{$\BU = \{\Bu_m\}^{M}_{m=1},~\Bu_m=\{u_m^l\}^L_{l=1}$}\;
		\mbox{$\BV = \{\Bv_m\}^{M}_{m=1},~\Bv_m=\{v_m^l\}^L_{l=1}$}\;
		\mbox{$\bm{\mu} = \{\mu^{l}\}^L_{l=1},\bm{\sigma} = \{\sigma^{l}\}^{L}_{l=1}$}\;
		\Return:\mbox{$~\CF_{\TT},\BU,\BV,\BK,\bm{\mu},\bm{\sigma}$.}
	}
\end{algorithm}

\subsection{Encrypt} \label{Encrypt}

We now propose an encryption method called DPRM to encrypt the dominated set $\HT$, so as to protect the model $\CF_{\BT}$ in terms of maximizing performance degradation. We run the DPRM independently on each subset $\HT_m$ of $\HT$ to facilitate the manipulation of decrypted parameters. The DPRM works by first generating a pseudorandom number sequence to mask $\HT_m$, and then transforming the masked version of $\HT_m$ into the same distribution as $\HT_m$. The detailed procedure of running DPRM is listed in Algorithm \ref{Alg:Encrypt}. For each $\HT_m$ in the partition of $\HT$, the DPRM sequentially performs:

{$\bm{1)}$} $RandMask(\HT_m,\kappa_m) \rightarrow \BC_{m}$: Given a secret key $\kappa_m$ produced by a key generator, we utilize the \emph{forward secure pseudorandom number generator} (FSPRGN) \cite{FSPRNG} $\mathcal{G}_{fs}$ to produce a pseudorandom number sequence $\BR_{m}$ with length $|\HT_m|$. The subset $\HT_m$ of the dominated set $\HT$ then is masked by this pseudorandom number sequence as $\BC_m = \HT_m + \BR_{m}$ (line 2).

{$\bm{2)}$} $Mapping(\BC_{m}) \rightarrow {\HC_{m}}$: An obvious limitation of $RandMask$ is that masked parameters $\BC_m$ are statistically different from the original $\HT_m$. Attackers can then easily identify $\BC_m$, which implies that locations of encrypted parameters are leaked. To tackle this serious threat, we have to further transform $\BC_m$ to another domain whose distribution is consistent with $\HT_m$. To this end, the fundamental issue now is to estimate the distribution of parameters in $\HT_m$.

Observing the histograms of parameters of several layers in the CNN model VGG19 \cite{VGG} in Fig. \ref{Fig:hist}, it is reasonable to infer that parameters in each layer of a CNN model follow symmetric distributions such as Gaussian. Hence, we assume that parameters $\BT^l$ of the $l$-th layer follow a Gaussian distribution $\mathcal{N}(\theta|\mu^l,\sigma^l)$, where the mean $\mu^l$ and standard deviation $\sigma^l$ can be estimated with sample mean and sample standard deviation of parameters in $\BT^l$.

Note that the masked parameters in $\BC_{m}$ are distributed across $L$ layers, where masked parameters in the $l$-th layer are denoted by $\BC^l_m$ ($l=1,...,L$). Making the distribution of $\BC_{m}$ be consistent with that of $\HT_m$ reduces to transform each $\BC^l_m$ into the Gaussian distribution $\mathcal{N}(\theta|\mu^l,\sigma^l)$ as $\HT^l \subset \BT^l$. As shown in Algorithm \ref{Alg:Encrypt} (lines 3-10), for $l=1,...,L$, we first scale elements in $\BC_{m}^{l}$ into the interval $[0,1]$, and then transform these scaled $c$'s with the inverse Gaussian CDF as follows:
\begin{equation} \label{eq:Scale}
\begin{small}
\begin{aligned}
& c \leftarrow (c-u^l_m)/(v^l_m-u^l_m), \quad \forall c \in \BC_{m}^{l}, \\
& \hat c \leftarrow F^{-1}(c| \mu^l, \sigma^l),
\end{aligned}
\end{small}
\end{equation}
where $u^l_m = \min (\BC^l_{m})$, $v^l_m = \max(\BC^l_{m})$, and $F^{-1}(\cdot| \mu^l, \sigma^l)$ is the inverse CDF of $\mathcal{N}(\theta|\mu^l,\sigma^l)$. After transforming all subsets $\BC_{m}^{l}$'s in $\BC_{m}$ as (\ref{eq:Scale}), we obtain the ciphertext $\HC_{m}$ of $\HT_m$ .
	
By repeatedly implementing the DPRM above on each $\HT_m$ in $\HT$, we encrypt the dominated set $\HT$ of $\CF_{\BT}$ into the ciphertext $\HC$. The protected model $\CF_{\TT}$ thus could be constructed by replacing $\HT$ in $\CF_{\BT}$ with $\HC$ (line 11). Apart from the protected model $\CF_{\TT}$, the {\bf{Encrypt}} module also generates secret keys $\BK$, scaling parameters $\BU, \BV$, and the statistical information $\bm{\mu}, \bm{\sigma}$, for generating permissions.

\begin{algorithm}[!t]
	\SetAlgoNoEnd
	\footnotesize{
	\caption{The scheme of $Decrypt$ module} \label{Alg:Decrypt}
	\KwIn{$\BS_{\Hm} = \{\BE_m, \Bu_m, \Bv_m, \kappa_m,\bm{\mu},\bm{\sigma}\}_{m=1}^{\Hm}$, protected model $\CF_{\TT}$.}
	\KwOut{Decrypted model: $\CF_{\BAT_{\Hm}}$}
	\For{$m = 1,...,\Hm$}{
		\mbox{$\HC_{m} \leftarrow \text{parameters~in~} {\CF}_{\TT} \text{~at~location~} \BE_m$}\;	
		\For{$l = 1,...,L$}
		{
			\mbox{$\BC^l_m = \emptyset, \HC_m = \{\HC^l_m\}^L_{l=1}$} \;
			\For{$\hat{c} ~ \mathbf{in}~\HC^l_m$}
			{
				\mbox{$c \leftarrow F(c| \mu^l, \sigma^l),~ c \leftarrow (\hc+u^l_m)(v^l_m-u^l_m)$.}
				\mbox{\footnotesize{$\rhd~\mu^l \in \bm{\mu},~\sigma^l\in\bm{\sigma},u^l_m\in\Bu_m,v^l_m\in\Bv_m$}}\;
				\mbox{$\BC^l_m \leftarrow \BC^l_m \bigcup \{c\}$}\;
			} 	
	}	
			\mbox{$\BC_m = \{\BC^l_m\}^L_{l=1}, \BR_{m} \leftarrow \mathcal{G}_{fs}(\kappa_m,|\BC_{m}|)$}\;
			\mbox{$\HT_m = \BC_m-\BR_{m}$}\;
	}
	\mbox{$\BAT_{\Hm} = \{\HT_{m}\}_{m=1}^{\Hm} \cup (\TT \slash \{\HC_{m}\}_{m=1}^{\Hm})$}\;
	\mbox{$\CF_{\BAT_{\Hm}}\leftarrow ~\text{replace}~ \TT ~\text{of}~ \CF_{\TT} ~\text{with}~ \BAT_{\Hm}$}\;
	\Return{$\CF_{\BAT_{\Hm}}$}}
\end{algorithm}

\subsection {Assign} \label{DA}

The performance of the protected model $\CF_{\TT}$ now is heavily suppressed since dominated parameters $\HT$ are encrypted. Users could access $\CF_{\TT}$ only with permissions composed of locations $\BE$ of dominated parameters and other five security-related components outputted from the {\bf{Encrypt}} module. More precisely, let $\BS_{\Hm}$'s $({\Hm}=1,...,M)$ be $M$ different levels of permissions, each of which is generated as follows:
\begin{equation}
\begin{small}
\BS_{\Hm} = \{\BE_m,\Bu_m,\Bv_m,\kappa_m, \bm{\mu}, \bm{\sigma} \}_{m=1}^{\Hm}.
\end{small}
\end{equation}
Here, each item $\{\BE_m,\Bu_m,\Bv_m,\kappa_m, \bm{\mu}, \bm{\sigma}  \}$ could independently decrypt the ciphertext of the dominated parameter subset $\HT_m$ in $\HT$. Details will be introduced in Section \ref{Sec:Decrypt}.

Also, the size of a permission $\BS_{\Hm}$ in bits is $(b_{\kappa}+64L)\Hm+\frac{16L\Hm}{M}\phi$ (please refer to supplementary materials for detailed calculations), where $b_{\kappa}$ is the bit size of the secret key $\kappa$, and $\phi$ is the number of dominated parameters in each layer. For the prevailing models, VGG19 and DnCNN considered in our experiment, the size of permissions is less than 508KB. Such overhead is acceptable in practice.

\subsection {Decrypt}\label{Sec:Decrypt}	

When receiving a permission $\BS_{\Hm}$ from users, the {\bf{Decrypt}} module is operated as follows. For each item $\{\BE_m,\Bu_m,\Bv_m,\kappa_m, \bm{\mu}, \bm{\sigma} \}$ in $\BS_{\Hm}$, the decryption procedure starts with locating the ciphertext $\HC_m$ of dominated parameter subset $\HT_{m}$ according to the location $\BE_m$ (line 2). The decryption of $\HC_m$ is merely the inverse process of the encryption of $\HT_m$. As shown in Algorithm \ref{Alg:Decrypt}, for those encrypted parameters $\HC^l_m$ of $\HC_m$ in the $l$-th layer, we first transfer them back to the random masked version of $\HT^l_{m}$, i.e.,  $\BC^l_m$, via the following transformations (lines 6-7)
\begin{equation} \label{eq:IDP}
\begin{small}
\begin{aligned}
\BC^l_m = \{c|\forall \hc \in \HC^l_m, & c \leftarrow F(\hc| \mu^l, \sigma^l),\\
& c \leftarrow (c+u^l_m)(v^l_m-u^l_m) \},\\
\end{aligned}
\end{small}
\end{equation}
where $F(\cdot,\mu^l, \sigma^l)$ is the CDF of Gaussian distribution $\mathcal{N}(\theta|\mu^l,\sigma^l)$. By repeating the above procedure (\ref{eq:IDP}) for $l=1,...,L$, we obtain the masked version $\BC_m$ of dominated parameters subset $\HT_m$. Then, we input the key $\kappa_m$ to the FSPRNG to generate the same random number sequence $\BR_{m}$ used to mask $\HT_m$ (line 8). The subset $\HT_m$ readily is decrypted by removing $\BR_{m}$ from $\BC_m$, i.e., $\HT_m = \HC_m-\BR_{m}$ (line 9).

With the permission $\BS_{\Hm}$, we can recover $\Hm$ dominated parameter subsets $\{\HT_{m}\}_{m=1}^{\Hm}$ in $\HT$ by repeating the above procedure.

The resulting decrypted model is denoted by $\CF_{\BAT_{\Hm}}$ (lines 10-11). Obviously, when $\Hm<M$, $M-\Hm$ dominated parameter subsets $\{\HT_{m}\}_{m=\Hm+1}^M$ are still encrypted in $\CF_{\BAT_{\Hm}}$. Hence it is a partially encrypted version of $\CF_{\BT}$. As $\Hm$ increases to $M$, all dominated parameters $\HT = \{\HT_{m}\}_{m=1}^M$ are recovered, and $\CF_{\BAT_{M}} = \CF_{\BT}$. Thus, the full performance of $\CF_{\BT}$ is then released. On the contrary, if no permissions are fed into the {\bf{Decrypt}} module, the performance of $\CF_{\BT}$ is still suppressed in the encrypted model. Consequently, we have provided users with hierarchical services (different performance levels) of the pretrained model and achieved access control against unauthorized users.

\begin{figure}[!t]
	\centering
	\includegraphics[width=0.62\linewidth]{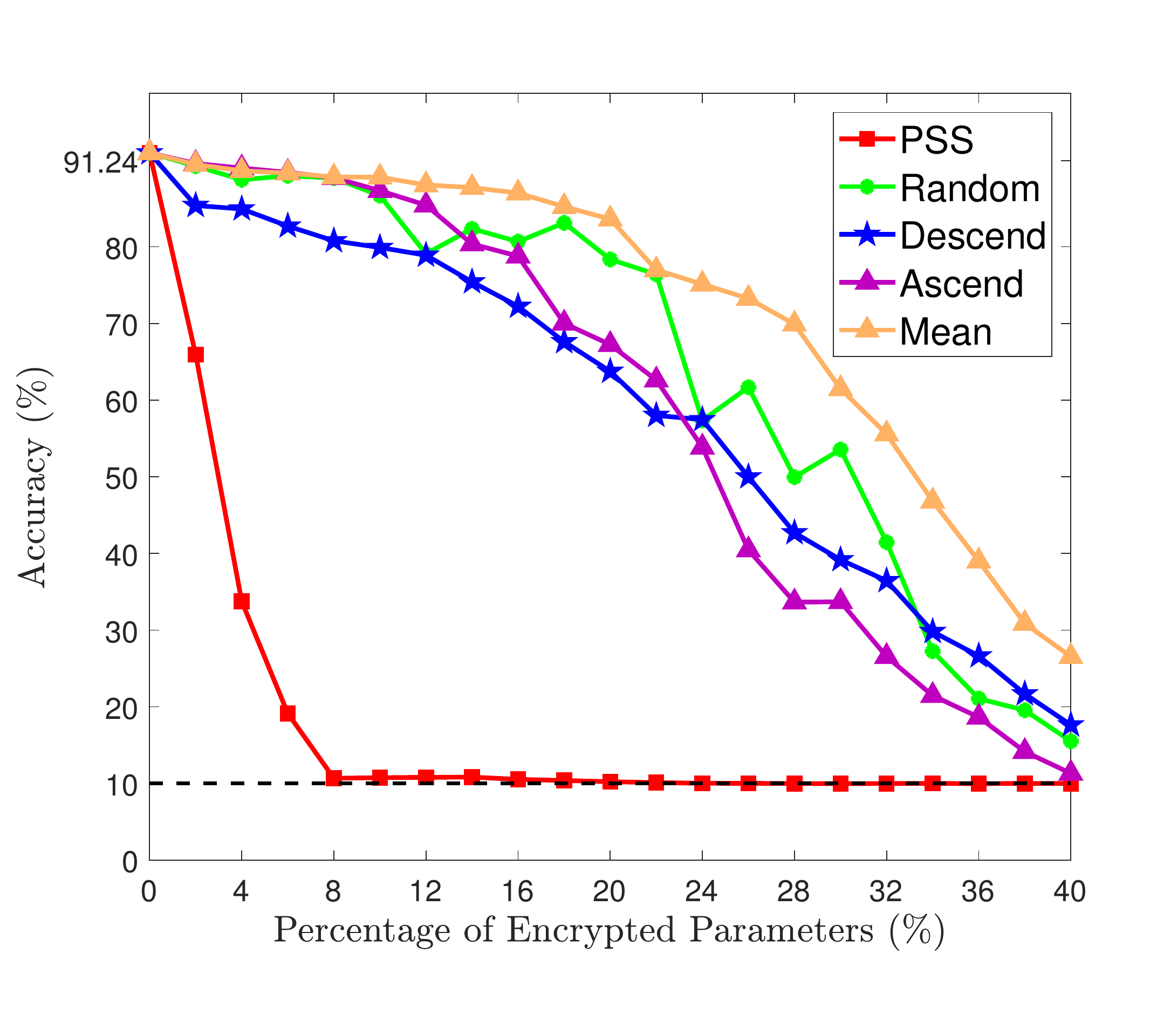}
	\caption{\small{The classification accuracy of the protected VGG19 on CIFAR10 with respect to different percentages of encrypted parameters.}}
	\label{Fig:VGG}
	\vspace*{-0.1in}
\end{figure}

\section{Imperceptibility of ciphertext} \label{Sec:Imperceptibility}

In this section, we theoretically prove the imperceptibility of the ciphertext, which significantly enhances the security of the protected model.

Let $\TT^l$ and $\BT^l$ be parameters of the $l$-th layer of the protected model $\CF_{\TT}$ and the pretrained model $\CF_{\BT}$, respectively. Recall that $\TT^l$ is a partially encrypted version of $\BT^l$ where dominated parameters $\HT^l$ are encrypted into the ciphertext $\HC^l$. To ensure that attackers cannot recover these dominated parameters $\BT^l$ as discussed in Section \ref{Sec:Threat}, a prerequisite is the imperceptibility of the ciphertext $\HC^l$. Taking $\TT^l$ as a noisy version of $\BT^l$, the dominated parameters $\HT^l$ of $\BT^l$ are contaminated by the noise $\HW^l=\HC^l - \HT^l$. If attackers cannot perceive the added noise $\HW^l$ by observing $\TT^l$, the ciphertext $\HC^l$ is also imperceptible. To prove that $\HW^l$ is imperceptible, we resort to the measure equivocation defined by Shannon \cite{shannon}, which evaluates the information leakage of $\HW^l$ when observing $\TT^l$.  This measure is also widely used in the field of steganography and watermarking \cite{eco2,eco1}. In our analysis, we modify the original equivocation into an equivalent form, $\hat{E}(\HW^l,\TT^l) = {I(\HW^l;\TT^l)} / H(\HW^l)$, where $H(\HW^l)$ is the entropy of $\HW^l$ and $I(\HW^l;\TT^l)$ is the mutual information between $\HW^l$ and $\TT^l$. If we can prove that $\hat{E}(\HW^l,\TT^l)$ is negligible, the information leakage of $\HW^l$ is also negligible, and thus $\HW^l$ is imperceptible. We have the following theorem on the magnitude of $\hat{E}(\HW^l,\TT^l)$.

\begin{theorem} \label{THM:eoc}
	The modified equivocation between $\HW^l$ and $\TT^l$ is of order $|\HW^l|^{-1/2}$. That is, as $|\HW^l|\rightarrow \infty$,
	\begin{equation} \label{eq:leakage}
	\begin{small}
	\hat{E}(\HW^l,\TT^l) = O(|\HW^l|^{-1/2}).
	\end{small}
	\end{equation}  	
\end{theorem}

The detailed proof is given in the supplementary materials. As can be easily seen from (\ref{eq:leakage}), $\hat{E}(\HW^l,\TT^l)$ is negligible if $|\HW^l|$ is large enough, or equivalently if the number of dominated parameters in $\HT^l$ is large enough. This condition is quite reasonable in practice as CNN models typically contain a massive of parameters. Taking VGG19 as an example, the number of parameters in the 8-th layer is of order $10^{6}$. Supposing $10\%$ parameters are identified as dominated ones, the order of $|\HW^l|$ would be $10^{5}$ and that of $\hat{E}(\HW^l,\TT^l)$ is $10^{-5/2}$.  Thus we claim that the ciphertext in the protected model is imperceptible.

\begin{figure}[!t]
	\centering
	\includegraphics[width=0.62\linewidth]{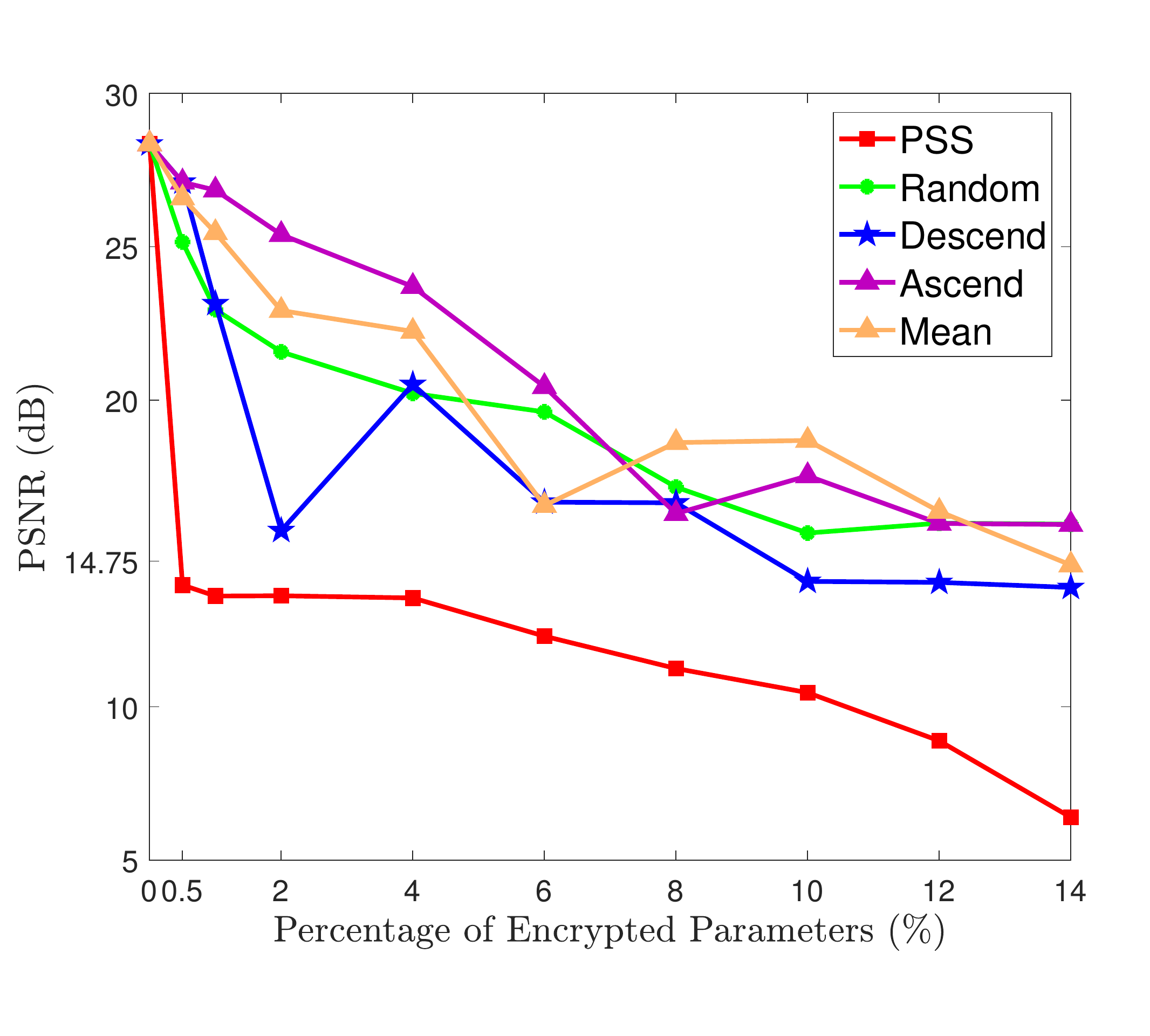}
	\caption{\small{The denoising performance of the protected DnCNN with respect to different percentages of encrypted parameters.}}  \label{Fig:DnCNN}
	\vspace*{-0.1in}
\end{figure}

\section{Experiments results}\label{Sec:Exps}

In this section, we experimentally verify that our system can achieve the goals defined in Section \ref{Sec:goals}. At first, we briefly introduce the experiment setup.

We consider two CNN models: the classification model VGG19 \cite{VGG} and the denoising model DnCNN \cite{DnCNN}. We train VGG19 on CIFAR10 \cite{Cifar10} and DnCNN on 300 noisy images from ImageNet \cite{Imagenet} with Gaussian noise (noise level 50). The best classification accuracy of the pretrained VGG19 on 10000 test images of CIFAR10 is $91.24\%$, and the best PSNR of the pretrained DnCNN on 40 noisy images is 28.35dB. According to our observations, encrypting several layers of VGG19 and DnCNN is enough to cause the maximal performance degradation. Therefore, in all experiments below, we only selectively encrypt parameters of 1-st, 2-th, 5-th, and 9-th layers of VGG19, and that of DnCNN are 6-th, 9-th, and 12-th layers. Due to the space limit, more experiment results can be found in the supplementary materials.

\begin{table}[!t]
	\centering
	\scalebox{0.8}{
		\begin{tabular}{c c c c c c c}
			\toprule
			\multirow{2}{*}{\vspace*{-0.05in}Model}&\multicolumn{6}{c}{Level of Permissions: $\Hm$} \\
			\cmidrule[0.5pt](lr){2-7}
			&0&1&2&3&4&5 \\
			\midrule
			VGG19 ($\%$) &10.00&65.41&78.65&82.58&87.33&91.24 \\
			DnCNN (dB) 	 &13.61&20.12&23.23&25.52&26.58&28.35 \\
			\bottomrule
	\end{tabular}}%
	\caption{Hierarchical Performance of Decrypted Models}
	\label{tab:Hierarchy}%
	\vspace*{-0.2in}
\end{table}%

\subsection{Effectiveness of the proposed SE}\label{Effectiveness}	

For rejecting unauthorized access, a successful protection should degrade the performance of the encrypted model into the worst situation, if no permission is granted. For instance, for VGG19 on CIFAR10, the worst prediction accuracy is $10\%$ (random guess among 10 classes). To the best of our knowledge, our proposed scheme is the first one to protect CNN models with SE. For preparing the competing algorithms, we design four different strategies to select parameters from considered layers, and encrypt them with the proposed DPRM. 1) {$\mathbf{Random}$}: We select parameters uniformly at random; 2) {$\mathbf{Mean}$}: We extract parameters around the mean value. This selection strategy is motivated by the observation that parameters are concentrated around the mean value (see Fig. \ref{Fig:hist}); 3) {$\mathbf{Descending}$}: A reasonable hypothesis is that the importance of parameters is positively correlated to their values. Thus, we select parameters in the descending order of their values; and 4) {$\mathbf{Ascending}$}: Conversely, we select parameters in ascending order.

Fig. \ref{Fig:VGG} shows the classification accuracy of VGG after encrypting different percentages of parameters of considered layers. We can observe that our PSS (red curve) degrades the VGG19 to the worst case when only $8\%$ of parameters are encrypted, while the best competing one needs to encrypt $40\%$ of parameters. Such a result demonstrates the effectiveness of the proposed SE for protecting VGG19. Similarly, Fig. \ref{Fig:DnCNN} shows the denoising performance of the DnCNN when PSS and other competing algorithms are used for the parameter selection. As can be observed, the denoising performance degrades very quickly when model parameters are selected by PSS and encrypted. Note that, to eliminate the randomness, results in Fig. \ref{Fig:VGG} and \ref{Fig:DnCNN} are the averages over repeating the encryption for 20 times.

\begin{figure*}[!t]
	\centering
	\subfigure[Clean ]{\label{Fig:Ori} \includegraphics[width=0.14\textwidth]{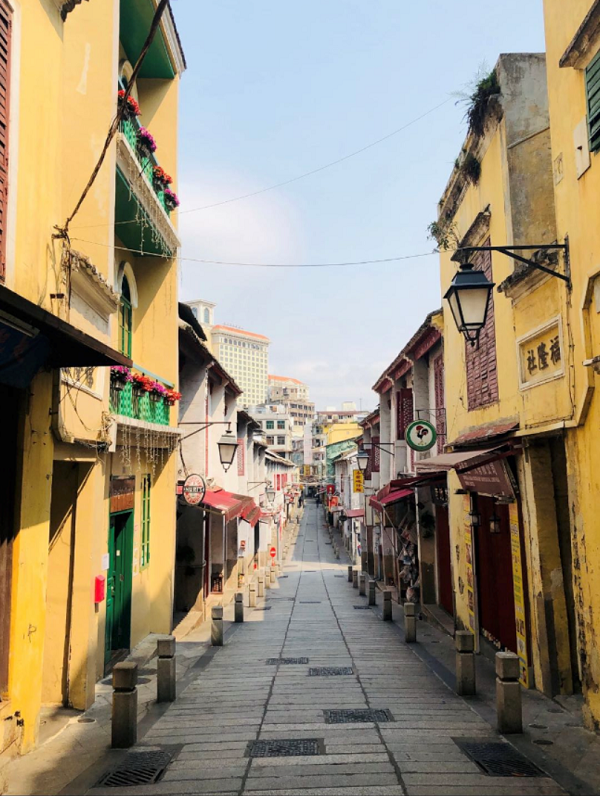} }
	\subfigure[Noisy (14.15dB)]{\label{Fig:Noisy} \includegraphics[width=0.14\textwidth]{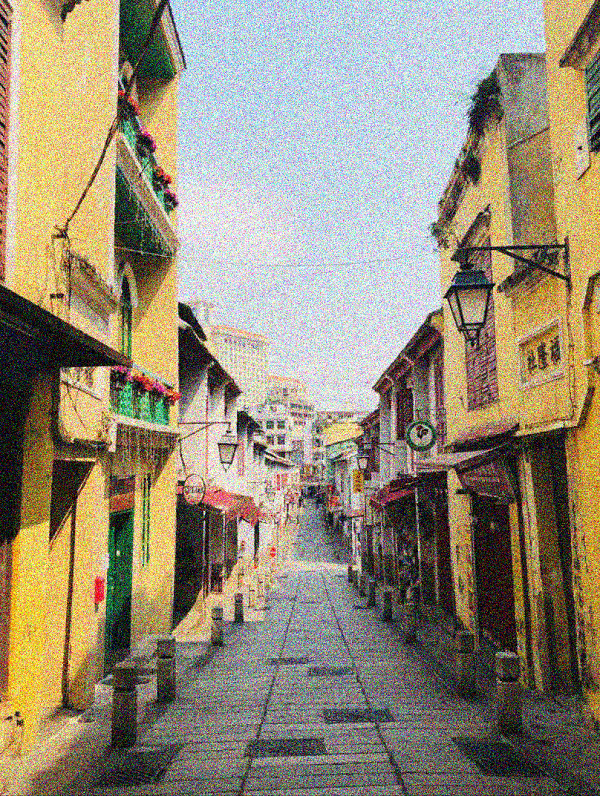}}
	\subfigure[$\Hm=0$ (8.91dB)]{\label{Fig:m0}  \includegraphics[width=0.14\textwidth]{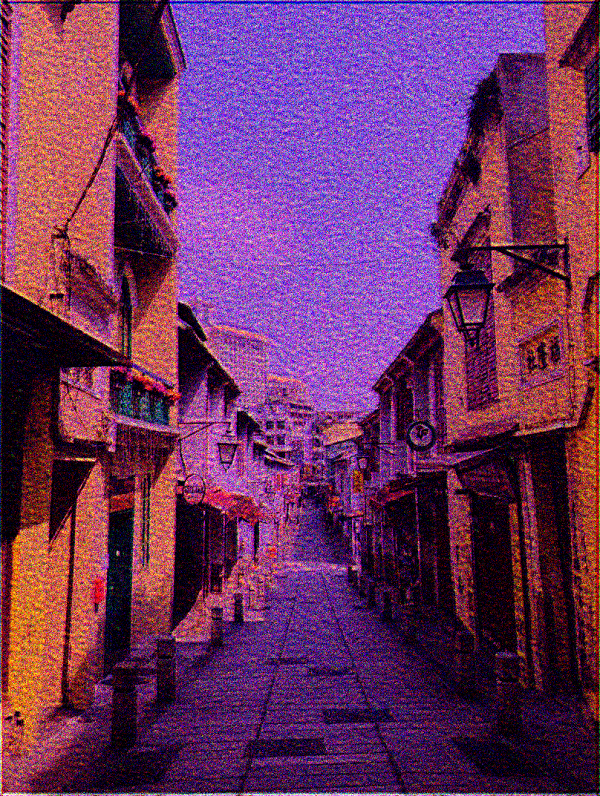}}
	\subfigure[$\Hm=1$ (19.22dB)]{\label{Fig:m1} \includegraphics[width=0.14\textwidth]{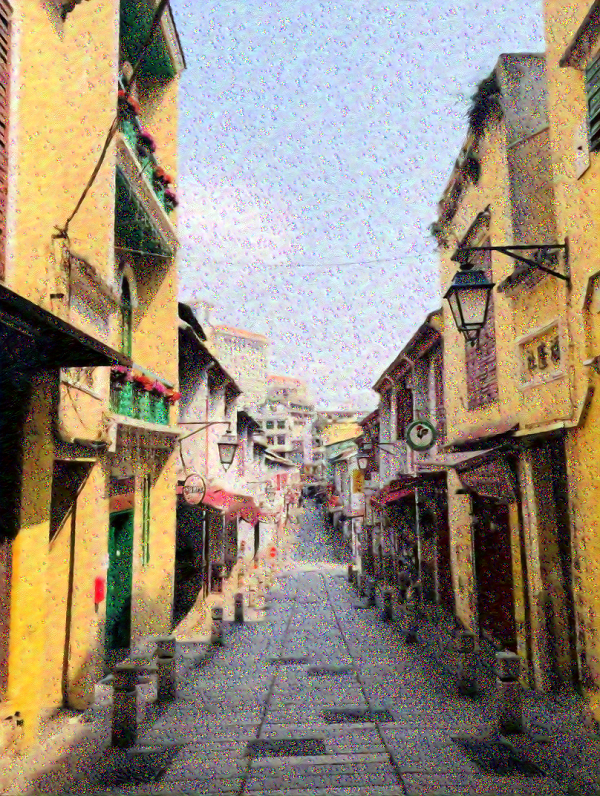} }
	\subfigure[$\Hm=3$ (24.40dB)]{\label{Fig:m4} \includegraphics[width=0.14\textwidth]{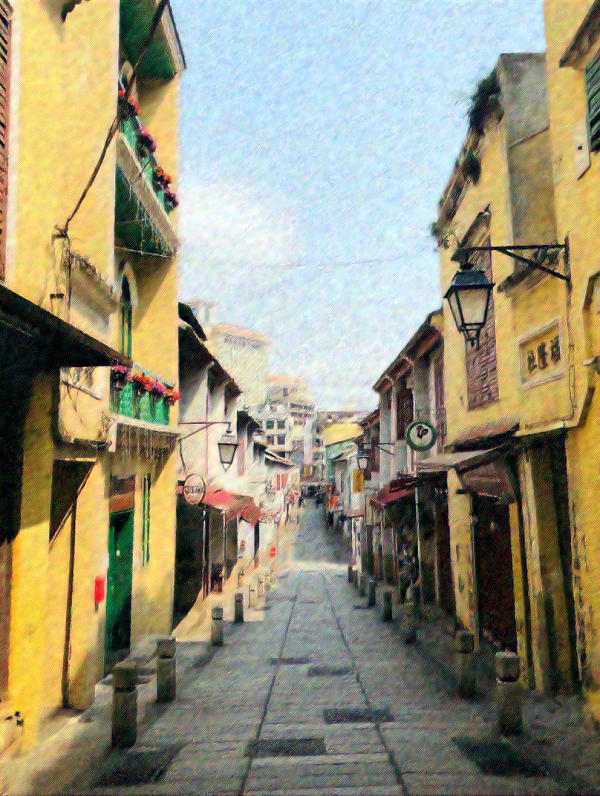}}
	\subfigure[$\Hm=5$ (29.91dB)]{\label{Fig:m6}  \includegraphics[width=0.14\textwidth]{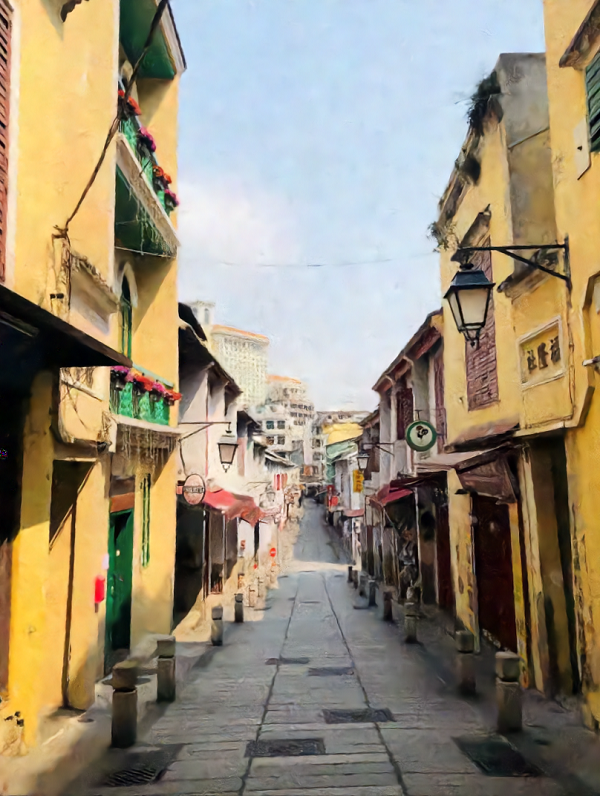}}
	\caption{\small{Images to illustrate the hierarchical performance of the protected DnCNN with different permissions. (a) Clean image; (b) Noisy image; (c) The output image of the protected DnCNN without permission; (d-f) Denoising results of the noisy image  by utilizing the protected DnCNN with different permissions $\BS_{\Hm}$'s.}}
	\label{Fig:Imgs}
	\vspace{-0.1in}
\end{figure*}

\subsection{Hierarchical performance of the released model}\label{Hierarchy}
	
We now demonstrate that models decrypted from the protected model with various permissions could exhibit different levels of performance. We selectively encrypt $10\%$ ($2\%$) parameters of considered layers of the VGG19 (DnCNN). Then, 5 permissions $\BS_{\Hm}$ ($\Hm = 1,...,5$) are generated by the module $\mathbf{Assign}$ in Section \ref{DA} ($M=5$). These permissions are fed into the $\mathbf{Decrypt}$ module to decrypt the protected VGG19 (DnCNN). The performance of the decrypted VGG19 and DnCNN with respect to the $5$ permissions is recorded in Table \ref{tab:Hierarchy}. As the increase of $\Hm$, the decrypted VGG19 (DnCNN) exhibits different levels of accuracy (PSNR) and reaches the best one when inputting the highest permission ($\Hm=5$). Here, $\Hm=0$ implies no permission is granted, corresponding to the worst prediction accuracy. For a better illustration of the hierarchical performance of the released DnCNN, the visualized denoising results of a test image under different permissions are shown in Fig. \ref{Fig:Imgs}. One can see from Fig. \ref{Fig:Imgs}(d)-(f) that a higher level of permission (larger $\Hm$) endows users with a better denoised image. Interestingly, for users without permission ($\Hm=0$, see Fig. \ref{Fig:m0}), the resulting denoised image is even worse than the original one.

\begin{table}[!t]
	\centering
	\scalebox{0.7}{
		\begin{tabular}{c c c c c c c }
			\toprule
			\multicolumn{7}{c}{Attacking Goals: \quad VGG19 = 65.41\% \quad  DnCNN = 20.12dB} \\
			\midrule
			\multirow{2}{*}{\vspace*{-0.05in}Model}&\multicolumn{3}{c}{Denoising~via~Wavelets}&\multicolumn{3}{c}{Denoising~via~Filters} \\
			\cmidrule[0.5pt](lr){2-7}
			&\bf{DB2}&\bf{Haar}&\bf{Sym9} &\bf{Average}&\bf{Gaussian}&\bf{Median} \\
			\midrule
			VGG19&21.15\%&23.13\%&24.72\%&19.54\%&18.42\%&17.31\% \\
			\midrule
			DnCNN&15.17dB&16.71dB&15.31dB&13.54dB&14.32dB&15.51dB \\
			\midrule
			\midrule
			Model&\multicolumn{3}{c}{Layer-wise}&\multicolumn{3}{c}{Transferring} \\
			\midrule
			VGG19&\multicolumn{3}{c}{55.15\%}&\multicolumn{3}{c}{39.14\%} \\
			\midrule
			DnCNN&\multicolumn{3}{c}{18.24dB}&\multicolumn{3}{c}{17.49dB}\\
			\bottomrule
	\end{tabular}}%
	\caption{Performance of Protected Models under Attacks}\label{Tab:CAttack}
	\vspace*{-0.2in}
	\label{tab:generiablity}%
\end{table}%

\subsection{Security against potential attacks}\label{Circumventing}	

We then evaluate the security of the protected model against denoising and retraining attacks considered in the threat model. First of all, we define the attackers' capability and goals to quantitatively evaluate whether an attack succeeds. $\mathbf{Attackers'~Capability}$: Attackers could challenge the protected model under one or all of the following capabilities. 1) Attackers can access all parameters of the protected model; 2) Attackers possess limited data ($10\%$) used for training the pretrained model; 3) Attackers own another model that has the same structure as the protected one; but is trained over another dataset; and 4) Attackers have known which layers in the model are encrypted. $\mathbf{Attackers'~Goal}$: The attackers' goal is to obtain the same model performance as a user with the lowest permission. As shown in Table \ref{tab:Hierarchy}, for VGG19 and DnCNN, the attackers' goals are 65.41\% classification accuracy and 20.12dB of the denoised image, respectively.

We now consider several potential denoising and retraining attacks and show that the protected model is secure against them. Note that, in Section \ref{Sec:Imperceptibility}, we have proved the imperceptibility of the encrypted parameters. Hence, all attacks discussed below are based on the premise that the locations of the encrypted parameters are unknown.

\underline{\textbf{DENOISING~ATTACKS}}

1) \textbf{Denoising via Wavelets}: Attackers take selectively encrypted parameters in each considered layer as a partially contaminated discrete signal (flattening parameter tensors into a 1-D vector). They can remove the noise by resorting to wavelet denoising techniques. To simulate attackers' behavior, we consider three different wavelets: $\mathbf{DB2}$, $\mathbf{Haar}$, and $\mathbf{Sym9}$. As shown in Table \ref{Tab:CAttack}, the accuracies of the protected model suffering the three wavelet based denoising attacks are  21.15\%, 23.13\%, and 24.72\%, respectively. Obviously, none of them achieves an accuracy better than the attacking goal 65.41\%. For DnCNN, the best performance given by the three types of attack is 16.71dB, which is still inferior to the attacking goal (20.12dB). Thus, we conclude that the VGG19 and DnCNN protected by our scheme are secure against the wavelets based denoising attacks.

2) \textbf{Denosing via Filters}: The significant performance degradation of the encrypted model possibly ascribes to some abnormally large or small parameters. Attacker could implement the average filter or the median filter on the 1-D signal of encrypted parameters. Moreover, the Gaussian filter possibly is suitable since the noise used to encrypt dominated parameters follows a Gaussian distribution. The performance of the protected VGG19 and DnCNN under these filter based attacks is recorded in Table \ref{Tab:CAttack}. As can be seen, for VGG19, all the restored accuracies after attacks cannot exceed the attacking goal (65.41\%). Similarly, for DnCNN, the best PSNR of the attacked model is 15.51dB, which is still worse than the attacking goal (20.12dB). Therefore, the encrypted VGG19 and DnCNN models are secure against the filtering based attacks.

\underline{\textbf{RETRAINING ATTACKS}}
	
1) \textbf{Layer-wise}: Since only several layers are encrypted by our proposed scheme, attackers could retrain each layer independently by fixing parameters of other layers, based on the available training data. The classification accuracy of the retrained VGG19 is recorded in Table \ref{Tab:CAttack}. It can be seen that the resulting accuracy is $55.15\%$, which is lower than the attacking goal 65.41\%. A similar result for DnCNN is 18.24dB, which is also worse than the attacking goal (20.12dB). Thus, under the assumption on attackers' capability and the defined attacking goal, we conclude that this type of retraining attack is not successful.

2) \textbf{Transfering}: Attackers may hypothesize that the distribution of encrypted parameters possibly is consistent among models with the same structure; but trained on another dataset. They thus implement the proposed PSS strategy on a new VGG19 trained on other ten classes from CIFAR100 \cite{Cifar10} and obtain locations of dominated parameters of this VGG19. Then, they retrain parameters of the protected VGG19 at locations learned from the new VGG19. The performance of such attacked model is only 39.14\%, far from the attacking goal 65.41\%. We also try to attack DnCNN by using the same strategy, where the locations of encrypted parameters are transferred from another DnCNN trained over the noisy images with noise level 25. In this case, the PSNR value of the denoised image is 17.49dB, which is still lower than the attacking goal (20.12dB).

\section{Conclusions}\label{Sec:Conclusion}
In this paper, we have proposed a SE algorithm to protect a CNN model by firstly selecting important parameters from this model with the PSS and then encrypting the selected parameters with DPRM. A system based on our SE can prevent a CNN model from unauthorized access and also provide authorized users with hierarchical services. Experimental results have been provided to show the effectiveness and security of the SE on protecting CNN models. {\bfseries{Acknowledgments:}} This work was supported by Macau Science and Technology Development Fund under SKL-IOTSC-2018-2020, 077/2018/A2, 0015/2019/AKP, and 0060/2019/A1, by Research Committee at University of Macau under MYRG2018-00029-FST and MYRG2019-00023-FST, by Natural Science Foundation of China under 61971476. This work was also supported by Alibaba Group through Alibaba Innovative Research Program.

{\small
\bibliographystyle{ieee_fullname}
\bibliography{myrefs}
}

\end{document}